\documentclass[11pt,a4paper]{article}

\usepackage[utf8]{inputenc}
\usepackage[T1]{fontenc}
\usepackage{amsmath,amssymb,amsthm}
\usepackage{graphicx}
\usepackage{booktabs}
\usepackage{tabularx}
\usepackage{hyperref}
\usepackage{xcolor}
\usepackage{tikz}
\usepackage{float}
\usepackage[numbers,sort&compress]{natbib}
\usepackage[margin=1in]{geometry}
\usepackage{enumitem}
\usepackage{url}

\pdfoutput=1

\title{Sentiment Analysis on Movie Reviews: A Deep Dive into Modern Techniques and Open Challenges}

\author{
Agnivo Gosai$^{1,\dagger}$, Shuvodeep De$^{2,\dagger}$, Karun Thankachan$^{3,\dagger}$, \\
Ramadan A. ZeinEldin$^{4}$, Ali Wagdy Mohamed$^{5}$, Seyed Jalaleddin Mousavirad$^{6,*}$ \\[2ex]
\small $^{1}$Independent Researcher, NY, USA; agnivo2007@gmail.com \\
\small $^{2}$Texas State University, San Marcos, TX, USA; vvg26@txstate.edu \\
\small $^{3}$Carnegie Mellon University, PA, USA; kthankac@alumni.cmu.edu \\
\small $^{4}$King Abdulaziz University, Jeddah, Saudi Arabia; rzainaldeen@kau.edu.sa \\
\small $^{5}$Zewail City of Science and Technology, Giza, Egypt \\
\small $^{6}$Mid Sweden University, Sundsvall, Sweden \\[1ex]
\small $^{\dagger}$These authors contributed equally \quad $^{*}$Corresponding author
}

\date{}

\begin{document}

\maketitle

\begin{abstract}
This paper presents a comprehensive survey of sentiment analysis methods for movie reviews, a benchmark task that has played a central role in advancing natural language processing. We review the evolution of techniques from early lexicon-based and classical machine learning approaches to modern deep learning architectures and large language models, covering widely used datasets such as IMDb, Rotten Tomatoes, and SST-2, and models ranging from Naive Bayes and support vector machines to LSTM networks, BERT, and attention-based transformers. Beyond summarizing prior work, this survey differentiates itself by offering a comparative, challenge-driven analysis of how these modeling paradigms address domain-specific issues such as sarcasm, negation, contextual ambiguity, and domain shift, which remain open problems in existing literature. Unlike earlier reviews that focus primarily on text-only pipelines, we also synthesize recent advances in multimodal sentiment analysis that integrate textual, audio, and visual cues from movie trailers and clips. In addition, we examine emerging concerns related to interpretability, fairness, and robustness that are often underexplored in prior surveys, and we outline future research directions including zero-shot and few-shot learning, hybrid symbolic--neural models, and real-time deployment considerations. Overall, this abstract provides a domain-focused roadmap that highlights both established solutions and unresolved challenges toward building more accurate, generalizable, and explainable sentiment analysis systems for movie review data.
\end{abstract}

\noindent\textbf{Keywords:} Sentiment analysis, movie reviews, opinion mining, natural language processing (NLP), text classification, machine learning, deep learning, emotion detection, feature extraction, polarity classification

\section{Introduction}
\subsection{Background}

Sentiment analysis of movie reviews, that is, the task of determining whether a piece of text conveys a positive, negative, or neutral opinion, is one of the most enduring and influential test beds in natural language processing (NLP)~\citep{pang2002thumbs,zhang2018deep}. What began a couple of decades ago as a seemingly straightforward binary classification problem has evolved into a sophisticated domain that exemplifies the full spectrum of challenges facing modern computational linguistics: from handling linguistic creativity and cultural nuance to ensuring robust deployment in resource-constrained environments~\citep{yadav2020sentiment,khurana2023natural}. The literature on movie review sentiment analysis is anchored by Pang and Lee's seminal 2008 monograph ``Opinion Mining and Sentiment Analysis'' in Foundations and Trends in Information Retrieval~\citep{article}, which established the definitive theoretical framework and extensively utilized movie review classification as the primary case study, building upon their pioneering 2002 EMNLP paper~\citep{pang2002thumbs} that first applied machine learning techniques to movie review sentiment classification. This foundational work, along with Maas et al.'s 2011 ACL paper~\citep{maas2011learning} introducing the Large Movie Review Dataset (IMDB), created the benchmark standards that continue to dominate the field today.

Recent comprehensive surveys (from 2020--2025) have documented the methodological evolution toward deep learning approaches, including Wankhade et al.~\citep{wankhade2022survey} and Jain et al.~\citep{jain2022systematic}, which provide extensive coverage of neural architectures and transformer models, while Raghunathan et al.~\citep{10176115} offers systematic literature review methodology focusing on contemporary challenges. Domain-specific surveys such as the conference paper by Tetteh et al. on sentiment analysis tools for movie review evaluation~\citep{10142834} and the recent publication examining the applications of BERT and XLNet~\citep{Danyal2024} to movie sentiment analysis demonstrate the field's continued focus on performance optimization using state-of-the-art models. However, existing reviews~\citep{birjali2021comprehensive, Dang2024Sentiment} on the analysis of movie review sentiment have notable gaps, particularly in addressing multilingual and cross-cultural contexts, bias and fairness, and the detection of sarcasm / irony. The coverage of aspect-based sentiment analysis is somewhat fragmented~\citep{Thet2010_AspectBasedMovieReviews, Horsa2023_AfaanOromooABSA, engproc2025087043}, with limited attention to methods targeting specific movie elements, and real-world deployment challenges are underexplored compared to academic benchmark performance. These omissions hinder a comprehensive understanding of the current state of the field and its practical applicability.

The evaluation landscape reveals inconsistent protocols and limited standardization across studies, with existing surveys~\citep{Bordoloi2023, MAO2024102048} failing to establish comprehensive frameworks for comparing methodological approaches. Recent surveys have also inadequately addressed emerging challenges including AI-generated fake reviews, cross-modal sentiment analysis that combines text with visual movie content, and privacy-preserving analysis techniques. These gaps indicate substantial opportunities for novel survey work that could provide more systematic coverage of multilingual approaches, comprehensive bias analysis, standardized evaluation frameworks, and practical deployment considerations that have been largely unexplored in the existing literature on this topic.

\subsection{Significance and Motivation}

Sentiment analysis of movie reviews has evolved from a niche academic pursuit into a critical infrastructure that today shapes the entertainment industry and influences billions of viewers around the world. In an era where digital platforms generate millions of user reviews daily, automated sentiment analysis has become the invisible but essential technology that powers recommendation systems on major streaming services such as Netflix, Amazon Prime and Disney+~\citep{poria2017review}. Studios use these systems to gauge audience reception before and after theatrical releases, marketing teams refine promotional strategies based on sentiment insights, and critics increasingly find their influence mediated through algorithmic aggregation systems~\citep{zhang2018deep} that must parse the subtle distinctions between professional critique and casual opinion. Movie reviews presented by audiences has an exceptionally rich linguistic landscape from the measured prose of professional critics to the enthusiastic hyperbole of fan communities, from the sardonic wit of satirical posts to the technical analysis of film school graduates~\citep{maas2011learning,socher2013recursive}, making this domain an ideal testbed for evaluating NLP systems across varied linguistic challenges, cultural contexts, and temporal periods~\citep{blitzer2007biographies}. As automated sentiment understanding increasingly shapes commercial decisions and cultural discussions, improving its accuracy, robustness, and interpretability becomes a practical necessity as well as a significant scientific goal.

However, despite achieving near-perfect performance on standard benchmarks~\citep{devlin2019bert}, real-world deployment of sentiment analysis systems continues to reveal fundamental limitations that transcend simple accuracy measurements. \textbf{Sarcasm and irony} remain particularly challenging~\citep{wilson2005recognizing}, as rhetorical expressions such as ``Great, another 3-hour snooze-fest'' fundamentally invert the relationship between surface-level sentiment indicators and actual meaning. \textbf{Domain and temporal drift} introduce ongoing stability issues when models confront new platforms, evolving slang, and cultural references absent from their training data~\citep{blitzer2007biographies}. \textbf{Long-form context processing} challenges arise when detailed reviews exceed standard transformer input limitations~\citep{vaswani2017attention}, while \textbf{explainability and bias concerns} become critical as these systems move from research prototypes to production deployments affecting real commercial outcomes~\citep{ribeiro2016should}. Finally, \textbf{resource efficiency} demands create fundamental trade-offs between model sophistication and computational constraints of mobile and edge environments~\citep{hu2022lora,dettmers2023qlora}.

These challenges are deeply interconnected rather than isolated: effective sarcasm detection often requires extensive context~\citep{wang2016attention}, which conflicts directly with efficiency constraints~\citep{hu2022lora}. Robust domain adaptation must continually account for changing cultural expressions~\citep{blitzer2007biographies}; and explainability methods must generalize across domains while operating under strict resource budgets~\citep{ribeiro2016should}. This interconnected nature suggests that future progress will depend on holistic approaches, ones that jointly address multiple challenges instead of optimizing individual components in isolation~\citep{rogers2020primer}. In the context of movie review analysis, nuanced language and diverse author voices around such integrated solutions are essential to deliver the reliable, interpretable, and efficient sentiment insights that underpin both commercial success and scholarly advancement.

\subsection{The Evolution of Sentiment Analysis}

The foundational work by Pang and Lee (2002)~\citep{pang2002thumbs} marked a pivotal moment in computational sentiment analysis, introducing the Movie Reviews (MR) corpus and demonstrating that machine learning techniques could significantly outperform simple lexicon-based sentiment scoring methods~\citep{turney2002thumbs}. Their work established movie review sentiment analysis as a benchmark problem and created the methodological framework that would guide research for the next two decades~\citep{pang2004sentimental}. The subsequent introduction of the IMDB-Large dataset~\citep{maas2011learning} with its 50,000 balanced reviews further standardized evaluation practices and enabled more robust comparison across approaches.

The field has since progressed through distinct methodological eras, each bringing significant advances in both techniques and performance. From early bag-of-words approaches~\citep{pang2002thumbs,mcdonald2007structured} achieving $\sim$82\% accuracy to static word embeddings~\citep{mikolov2013efficient,pennington2014glove} reaching 86--91\%, followed by RNN and attention mechanisms~\citep{hochreiter1997long,yang2016hierarchical} that reached 90--93\%, transformer architectures~\citep{vaswani2017attention,devlin2019bert,liu2019roberta} pushing performance to 94--97\%, and modern large language models~\citep{brown2020language,touvron2023llama} reaching 97--98\% on standard benchmarks~\citep{caiac2024sentiment}, this progression represents one of the most successful long-term research trajectories in NLP.

However, this apparent success reveals a more complex reality: as benchmark performance has approached theoretical limits, persistent challenges have emerged that cannot be solved through optimizing for better classification accuracy alone~\citep{ribeiro2016should,rogers2020primer}. Contemporary research has identified five core challenges that continue to define the frontier of movie review sentiment analysis~\citep{zhang2018deep,yadav2020sentiment}: sarcasm and irony detection, domain and temporal drift adaptation, long-form context processing, explainability and bias mitigation, and resource-efficient deployment. These challenges are not merely technical hurdles, but represent fundamental aspects of human communication that computational systems must learn to navigate.

\subsection{Limitations and Challenges}

Despite achieving near-human performance on benchmark datasets, current sentiment analysis methods encounter persistent limitations that prevent seamless real-world application. These limitations include:

\begin{itemize}[leftmargin=*]
\item \textbf{Sarcasm and Irony Detection}: Recognizing nuanced or indirect sentiment expressions remains problematic, significantly affecting accuracy.
\item \textbf{Domain and Temporal Drift}: Models trained on historical data or single-source reviews often fail when deployed in evolving or cross-platform contexts.
\item \textbf{Long-form Context Processing}: Contemporary models struggle to consistently interpret sentiment across extended texts, missing context-dependent meanings.
\item \textbf{Interpretability and Bias Mitigation}: Sentiment predictions by deep learning models frequently lack transparency, undermining trust in real-world deployments.
\item \textbf{Resource Efficiency}: High-performing transformer models require computational resources, limiting practical deployment scenarios, especially in real-time settings.
\end{itemize}

Thus, addressing these intertwined challenges is essential to bridge the gap between theoretical model performance and practical usability in the dynamic domain of movie review sentiment analysis.

\subsection{Novelty and Scope of this Review}

This review systematically synthesizes two decades of research from foundational lexicon-based approaches to cutting-edge transformer and multimodal models, uniquely emphasizing practical deployment considerations. Its primary contributions are:

\begin{itemize}[leftmargin=*]
\item \textbf{Comprehensive Methodological Evolution}: A structured analysis of sentiment analysis methodologies from early rule-based systems to modern large language models, contextualizing their strengths, limitations, and practical implications.
\item \textbf{Critical Analysis of Persistent Challenges}: Explicitly identifying and evaluating ongoing research limitations: sarcasm detection, domain adaptation, contextual understanding, interpretability, and computational efficiency, which previous reviews have only addressed separately or superficially.
\item \textbf{Integrated Benchmark Critique and Recommendations}: Proposing next-generation evaluation frameworks that incorporate multidimensional, real-world performance metrics, including temporal robustness, cross-platform adaptability, multimodal understanding, and resource-aware assessments.
\item \textbf{Future-Oriented Research Roadmap}: Highlighting promising future directions in multimodal fusion, zero-shot and hybrid model training, interactive explainability, and efficient deployment, which collectively address the identified limitations in an integrative manner.
\end{itemize}

We provide a structured analysis of movie review sentiment analysis that extends beyond traditional accuracy-focused evaluation to address the complex challenges facing modern deployment scenarios. We synthesize the complete methodological evolution of the field while maintaining focus on practical applicability and real-world deployment considerations, with particular emphasis on the five persistent challenges that define the current research frontier: sarcasm and irony detection, domain and temporal drift, long-form context processing, explainability and bias mitigation, and resource-efficient deployment.

\begin{figure}[htbp]
    \centering
    \includegraphics[width=\textwidth]{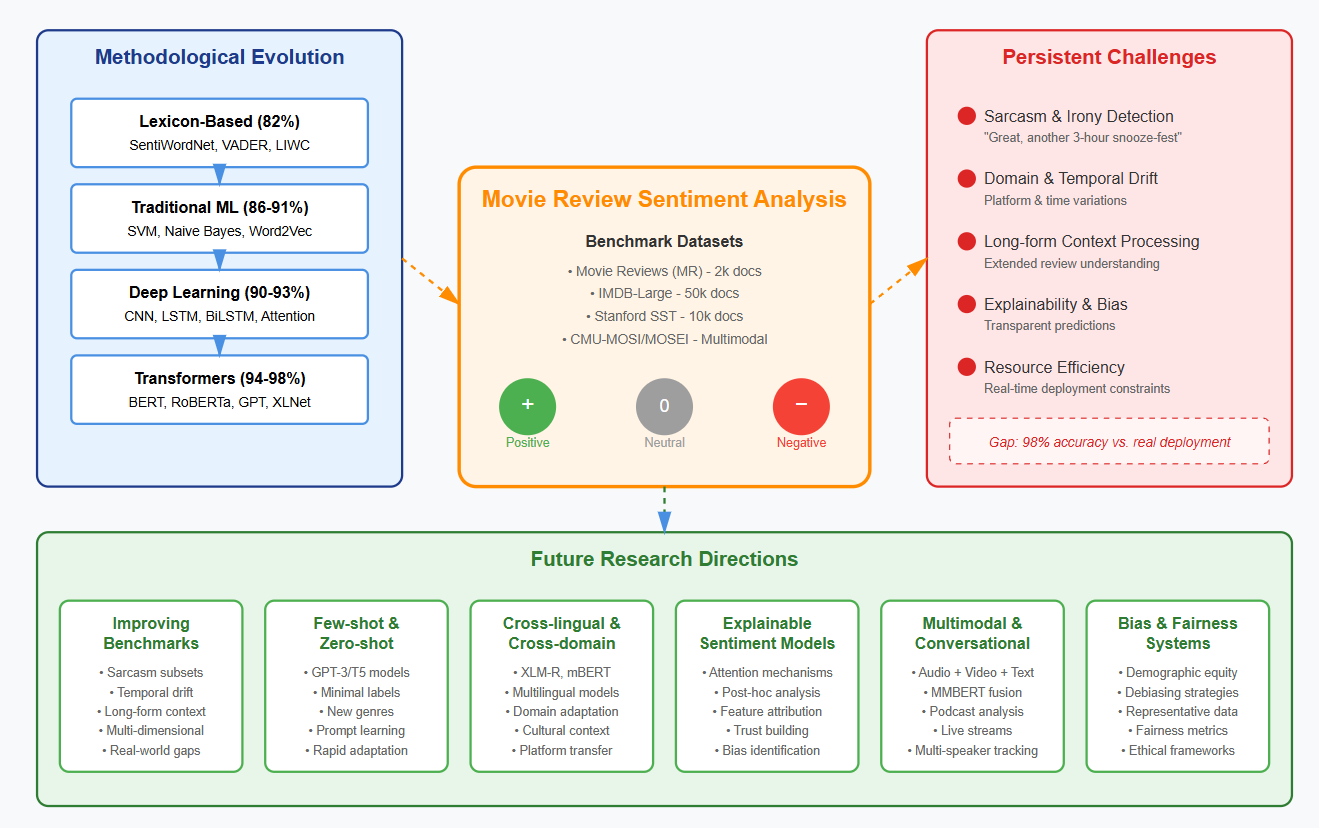}
    \caption{Sentiment Analysis on Movie Review: Organization of the Article. Figure created with AI assistance.}
    \label{fig:sentiment-framework}
\end{figure}

Our analysis is organized into seven complementary sections that build systematically from foundational concepts to cutting-edge research directions. \textbf{Section 2} establishes the theoretical foundation by examining different granularities of sentiment analysis and categorizing types of movie review data. \textbf{Section 3} traces the evolution of evaluation frameworks from foundational datasets all to multimodal sentiment extensions, critically analyzing how benchmarks have shaped research while potentially constraining real-world understanding. \textbf{Section 4} provides comprehensive coverage of the methodological progression from rule-based approaches to transformer models, documenting how it evolved from 82 percent accuracy to 97--98 percent accuracy. \textbf{Section 5} provides a detailed examination of the persistent challenges that define the current research in this area, highlighting how these interconnected problems require holistic solutions. \textbf{Section 6} explores some of the most promising research avenues, including improved benchmarks, few-shot learning, cross-lingual transfer, explainable models, multimodal analysis, and bias mitigation, while \textbf{Section 7} synthesizes key insights to develop robust and deployable systems. Throughout this analysis, we maintain focus on practical implications of research developments, bridging the gap between academic research and real-world deployment to serve both as a comprehensive resource for researchers and a strategic guide for practitioners. This field currently stands at a critical juncture where traditional accuracy-focused approaches have reached their limits, and this review provides the foundation for navigating the transition toward systems that can truly understand, explain, and generalize across the rich diversity of human expression in film criticism.

\section{Problem Formulation and Taxonomy}

Sentiment analysis is often referred to as opinion mining since it is essentially the computational study of people's opinions, attitudes, and emotions expressed in text. At its core, sentiment analysis seeks to infer the underlying sentiment polarity: positive, negative, or neutral conveyed by a writer about an entity or event. Early work in this field framed the problem as a binary or multiclass text classification task, leveraging hand-crafted lexicons or shallow statistical features to assign a polarity label to an entire document or sentence~\citep{pang2002thumbs}. As the field matured, researchers broadened this scope to capture richer sentiment dimensions, such as intensity (e.g., ``very happy'' vs. ``slightly happy''), emotion categories (e.g., joy, anger, sadness), and even finer-grained opinions towards specific aspects of an entity~\citep{liu2012sentiment}.

\subsection{Granularity of Sentiment Analysis}

Sentiment analysis can be organized along several levels of granularity. In \textbf{document-level} analysis, the goal is to predict the overall sentiment of a full review or article. This is often the simplest formulation, but may obscure conflicting sentiments directed at different aspects (for example, praising acting but criticizing the plot). \textbf{Sentence-level} analysis addresses this by assigning sentiment labels to individual sentences, thereby allowing more localized sentiment detection within longer texts. \textbf{Aspect-level} (or feature-level) sentiment analysis goes a step further by first identifying specific attributes of entity (such as ``acting,'' ``direction,'' or ``cinematography'' in movie reviews) and then determining the sentiment expressed toward each attribute. This finer granularity supports nuanced insights, such as simultaneously capturing praise for special effects and disappointment with story pacing~\citep{socher2013recursive}.

\subsection{Types of Movie Review Data}

Movie reviews come in diverse formats and levels of annotation, each providing different signals for sentiment analysis. In this subsection, we distinguish three principal types of movie review data that have driven research in the field.

\subsubsection{Unstructured Textual Reviews}

The most prevalent form of movie review data consists of free-form text written by users or professional critics. These unstructured reviews appear on platforms such as IMDb and Rotten Tomatoes, where authors express nuanced opinions, personal anecdotes, and rhetorical devices that convey their affective stance~\citep{pang2002thumbs}. Unlike short social-media posts, movie reviews often span multiple paragraphs and include both explicit sentiment (``I loved the stunning visuals'') and implicit cues (``It took me a half hour to recover from that ending''). The richness of this text makes natural language processing both challenging and rewarding, as models must learn to parse context, resolve coreference, and interpret discourse structure to accurately gauge sentiment intensity and direction.

\subsubsection{Structured Ratings and Metadata}

Complementing unstructured reviews are structured signals such as numerical star ratings (e.g., 1--10) and categorical labels (e.g., `Fresh' vs. `Rotten' on Rotten Tomatoes). These coarse-grained annotations provide readily available polarity information that can serve as distant supervision for training classifiers~\citep{liu2012sentiment}. In many datasets, each review is accompanied by metadata such as reviewer ID, timestamp, and genre labels, which enables temporal analysis and user-level modeling of sentiment trends. However, structured ratings often mask the rationale behind the score, motivating hybrid approaches that combine metadata with text to improve interpretability and performance.

\subsubsection{Annotated Corpora for Fine-Grained Analysis}

To capture sentiment at sub-sentential levels, researchers have developed richly annotated corpora. The Stanford Sentiment Treebank (SST) is a paradigmatic example: it provides phrase-level polarity labels across the parse tree of each sentence, allowing models to learn compositional sentiment patterns (e.g., `not good' vs. `good')~\citep{socher2013recursive}. Such corpora facilitate sentence-level sentiment classification as well as aspect-level analysis, where individual entities or attributes (like `acting,' `plot,' or `soundtrack') are annotated with their own sentiment tags. Even though such annotations are labor-intensive to produce, they have lead to significant advances in deep learning architectures, particularly recursive and attention-based models that can reason over tree structures and isolate sentiment-bearing subphrases.

\section{Benchmark Datasets and Metrics}

The evolution of movie review sentiment analysis has been fundamentally shaped by the benchmark datasets that define evaluation standards and research directions~\citep{maas2011learning,socher2013recursive}. While these datasets have enabled remarkable progress in classification accuracy, their design choices and inherent limitations have also constrained our understanding of real-world performance across the five persistent challenges identified in this review~\citep{pang2002thumbs,zhang2018deep}. This section examines the major benchmark corpora, their contributions to the field's development, and the emerging recognition that traditional evaluation frameworks may inadequately capture the complexity of deploying sentiment analysis systems in production environments.

\subsection{Foundational Datasets: Establishing the Benchmark Paradigm}

\subsubsection{The Movie Reviews Corpus: Pioneering Binary Classification}

The seminal work by Pang and Lee (2002)~\citep{pang2002thumbs} introduced the Movie Reviews (MR) corpus, which became the foundational benchmark for sentiment analysis research. This dataset, containing 2,000 movie reviews equally split between positive and negative sentiment, established several conventions that would influence the field for decades: binary classification as the primary task, balanced class distribution as the evaluation standard, and accuracy as the dominant performance metric.

The MR corpus demonstrated that machine learning approaches could significantly outperform lexicon-based methods, with Support Vector Machines achieving approximately 82\% accuracy using unigram features~\citep{pang2002thumbs}. This result established sentiment analysis as a viable machine learning problem and created the methodological framework that would guide subsequent research. However, the relatively small size of the dataset (2,000 documents) and its focus on professional reviews from a specific time period would later prove limiting for understanding model robustness across the challenges of sarcasm detection, domain adaptation, and temporal drift.

The balanced 50/50 polarity split, while useful for controlled evaluation, may not reflect the natural distribution of sentiments in real-world movie review platforms~\citep{maas2011learning}. This design choice, replicated across subsequent benchmarks, potentially oversimplifies the nuanced nature of movie criticism where mixed sentiments, qualified opinions, and contextual judgments are common~\citep{socher2013recursive}.

\subsubsection{Expansion and Standardization: The IMDB-Large Dataset}

The introduction of the IMDB-Large dataset by Maas et al. (2011)~\citep{maas2011learning} represented a significant leap forward in benchmark sophistication, providing 50,000 reviews split into 25,000 training and 25,000 test samples. This dataset became the de facto standard for movie review sentiment analysis, hosted on platforms like Kaggle and Stanford AI, and enabled more robust evaluation of deep learning approaches as they emerged throughout the 2010s.

The IMDB-Large dataset's larger scale allowed for more sophisticated model architectures and training procedures, supporting the transition from traditional machine learning to deep learning approaches~\citep{kim2014convolutional,yang2016hierarchical}. The dataset's availability and standardized train/test splits enabled meaningful comparison across different methodological approaches, contributing to the field's rapid progress from 86--91\% accuracy with static word embeddings to 94--97\% with transformer architectures~\citep{devlin2019bert,liu2019roberta}.

However, the IMDB-Large dataset also inherited and amplified certain limitations from its predecessor. The binary classification paradigm, while computationally convenient, may inadequately capture the complexity of sentiment expression in movie reviews~\citep{socher2013recursive}. Professional and amateur reviewers often express nuanced opinions that resist simple positive/negative categorization, particularly when discussing different aspects of films or comparing works within specific genres or cultural contexts.

\subsection{Diversification and Multimodal Extensions}

\subsubsection{Stanford Sentiment Treebank: Phrase-Level Granularity}

The Stanford Sentiment Treebank (SST)~\citep{socher2013recursive} introduced a different perspective on sentiment analysis with 10,000 documents focused on short snippets and phrase-level sentiment analysis tasks. This dataset provided fine-grained sentiment annotations at the phrase level, enabling models to learn compositional sentiment understanding rather than document-level classification alone.

The SST's contribution extends beyond simple scale to methodological sophistication, as it enables evaluation of models' ability to understand how sentiment composes across linguistic structures~\citep{socher2013recursive}. This capability is particularly relevant for handling complex rhetorical constructions common in movie reviews, where sentiment may shift or be qualified across different parts of a text. The phrase-level annotations provide insights into model behavior that document-level accuracy measurements cannot capture.

However, the focus on short snippets may limit the SST's ability to evaluate models on the long-form context processing challenge identified in this review. Movie reviews often develop arguments over extended passages, and phrase-level sentiment understanding, while valuable, may not capture the full complexity of document-level sentiment development~\citep{yang2016hierarchical}.

\subsubsection{Multimodal Sentiment: CMU-MOSI and MOSEI}

The emergence of multimodal datasets like CMU-MOSI and CMU-MOSEI~\citep{zadeh2017tensor,zadeh2018multi} brought additional complexity with 2,200--22,800 YouTube clips incorporating text, video, and audio sentiment analysis. These datasets convert 7-point sentiment scales to binary classifications, enabling comparison with traditional text-only benchmarks while introducing the possibility of multimodal sentiment understanding.

The multimodal datasets represent a significant expansion of the evaluation paradigm, acknowledging that movie reviews often reference visual and auditory elements of films that pure text analysis cannot capture~\citep{zadeh2017tensor}. A reviewer discussing ``haunting cinematography'' or ``jarring sound design'' makes references that could benefit from access to actual visual and audio content, suggesting directions for more comprehensive sentiment understanding.

However, the conversion from 7-point sentiment scales to binary classifications, while enabling comparison with traditional datasets, may lose important nuances in sentiment expression that are particularly relevant for movie reviews~\citep{zadeh2018multi}. The multimodal nature of these datasets also introduces additional complexity to evaluation, requiring metrics that can assess performance across text, video, and audio modalities simultaneously.

\subsubsection{Recent Benchmark Expansions: GoEmotions and TweetEval}

The year 2020 marked significant advances in sentiment benchmark development with the introduction of GoEmotions and TweetEval. GoEmotions~\citep{demszky2020goemotions}, developed by Google Research, introduced the largest manually annotated dataset for fine-grained emotion classification, comprising 58,000 Reddit comments labeled across 27 emotion categories plus neutral. Unlike previous datasets limited to basic sentiment polarity or six basic emotions, GoEmotions captures nuanced emotional states including admiration, amusement, curiosity, and disappointment, enabling more sophisticated emotion-aware applications. The dataset's taxonomy was developed in collaboration with psychologists, covering 12 positive, 11 negative, and 4 ambiguous emotion categories.

TweetEval~\citep{barbieri2020tweeteval} addressed the fragmentation problem in social media NLP by unifying seven heterogeneous Twitter-specific classification tasks including sentiment analysis, emotion recognition, irony detection, and hate speech detection into a standardized evaluation framework. With over 150,000 tweets and consistent train/validation/test splits, TweetEval enables meaningful comparison across models and has become the de facto benchmark for social media sentiment analysis. The accompanying Twitter-RoBERTa model, pre-trained on 58 million tweets, established strong baselines that continue to inform research on domain-specific language model adaptation.

These recent datasets reflect a broader trend toward more comprehensive evaluation frameworks that capture the complexity of real-world sentiment expression beyond simple binary classification.

\begin{table}[H]
\caption{Evolution of Major Sentiment Analysis Benchmark Datasets.}
\label{tab:benchmark_evolution}
\centering
\footnotesize
\setlength{\tabcolsep}{3pt}
\begin{tabular}{@{}lccclcp{2.8cm}@{}}
\toprule
\textbf{Dataset} & \textbf{Year} & \textbf{Size} & \textbf{Split} & \textbf{Source} & \textbf{Modality} & \textbf{Key Innovation} \\
\midrule
Movie Reviews (MR) & 2002 & 2k & 50/50 & Prof. reviews & Text & Binary classification \\
IMDB-Large & 2011 & 50k & 25k/25k & User reviews & Text & Large-scale evaluation \\
Stanford SST & 2013 & 10k & Variable & Prof. snippets & Text & Phrase-level granularity \\
Rotten Tomatoes & 2013 & 10k & 50/50 & Mixed & Text & Short snippet focus \\
CMU-MOSI & 2016 & 2.2k & 7pt$\to$bin & YouTube & T+V+A & Multimodal integration \\
CMU-MOSEI & 2018 & 22.8k & 7pt$\to$bin & YouTube & T+V+A & Large-scale multimodal \\
GoEmotions & 2020 & 58k & Train/Val/Test & Reddit & Text & 27 fine-grained emotions \\
TweetEval & 2020 & 150k+ & Standardized & Twitter & Text & Unified social media benchmark \\
\bottomrule
\end{tabular}
\end{table}

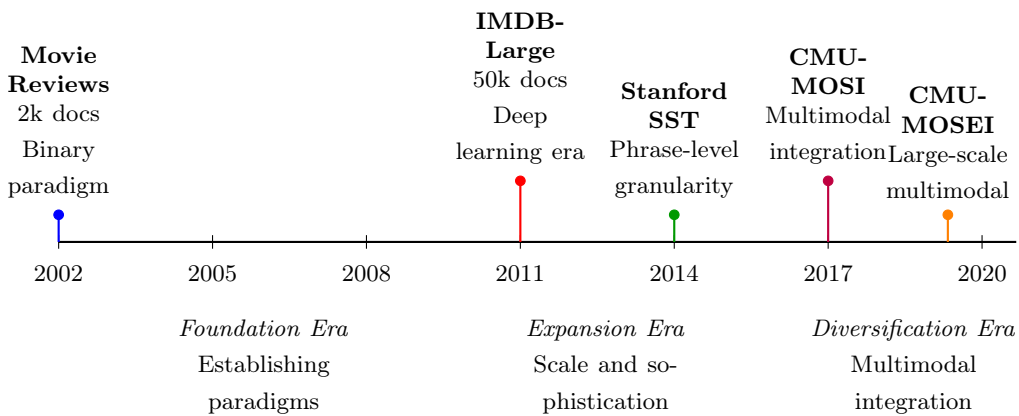
\begin{figure}[H]
\centering
\begin{tikzpicture}[scale=0.9]
\draw[thick] (0,0) -- (14,0);

\foreach \x/\year in {0/2002, 2.25/2005, 4.5/2008, 6.75/2011, 9/2014, 11.25/2017, 13.5/2020} {
    \draw (\x,-0.1) -- (\x,0.1);
    \node[below] at (\x,-0.2) {\footnotesize \year};
}

\node[above, text width=2cm, align=center] at (0,0.5) {\footnotesize \textbf{Movie Reviews}\\ 2k docs\\ Binary paradigm};
\draw[thick, blue] (0,0) -- (0,0.4);
\filldraw[blue] (0,0.4) circle (2pt);

\node[above, text width=2cm, align=center] at (6.75,1.0) {\footnotesize \textbf{IMDB-Large}\\ 50k docs\\ Deep learning era};
\draw[thick, red] (6.75,0) -- (6.75,0.9);
\filldraw[red] (6.75,0.9) circle (2pt);

\node[above, text width=2cm, align=center] at (9,0.5) {\footnotesize \textbf{Stanford SST}\\ Phrase-level\\ granularity};
\draw[thick, green!60!black] (9,0) -- (9,0.4);
\filldraw[green!60!black] (9,0.4) circle (2pt);

\node[above, text width=2cm, align=center] at (11.25,1.0) {\footnotesize \textbf{CMU-MOSI}\\ Multimodal\\ integration};
\draw[thick, purple] (11.25,0) -- (11.25,0.9);
\filldraw[purple] (11.25,0.9) circle (2pt);

\node[above, text width=2cm, align=center] at (13,0.5) {\footnotesize \textbf{CMU-MOSEI}\\ Large-scale\\ multimodal};
\draw[thick, orange] (13,0) -- (13,0.4);
\filldraw[orange] (13,0.4) circle (2pt);

\node[below, text width=3cm, align=center] at (3,-1) {\footnotesize \textit{Foundation Era}\\ Establishing paradigms};
\node[below, text width=3cm, align=center] at (8,-1) {\footnotesize \textit{Expansion Era}\\ Scale and sophistication};
\node[below, text width=3cm, align=center] at (12.5,-1) {\footnotesize \textit{Diversification Era}\\ Multimodal integration};

\end{tikzpicture}
\caption{Timeline of major benchmark dataset development in movie review sentiment analysis, showing the evolution from simple binary classification to complex multimodal evaluation frameworks.}
\label{fig:benchmark_timeline}
\end{figure}

\subsection{Evaluation Metrics}

\subsubsection{The Dominance of Accuracy and Macro-F1}

The standardization of evaluation metrics has been crucial for the field's development, with accuracy and macro-F1 remaining the dominant measures across most benchmark studies~\citep{zhang2018deep}. Many works additionally report ROC-AUC (Receiver Operating Characteristic - Area Under Curve) or MCC (Matthews Correlation Coefficient) to handle class imbalance issues, though the balanced nature of most movie review datasets makes this less critical than in other domains.

The choice of evaluation metrics reflects the field's evolution from simple binary classification to more nuanced understanding of model performance~\citep{yadav2020sentiment}. While accuracy provides an intuitive measure of overall performance, macro-F1 offers better insights into model behavior across both positive and negative classes. ROC-AUC provides valuable information about model discrimination ability across different threshold settings, which is particularly relevant for deployment scenarios where operating points may need adjustment.

However, these traditional metrics have significant limitations when evaluating models against the five persistent challenges identified in this review~\citep{ribeiro2016should}. Standard accuracy measurements may not capture a model's ability to handle sarcastic reviews, adapt to temporal drift, or maintain performance across different review lengths. This limitation has led to increased interest in developing more comprehensive evaluation frameworks that assess model robustness across multiple dimensions simultaneously.

\subsubsection{The Limitations of Binary Classification Paradigms}

The focus on binary classification in most benchmarks, including the conversion of multi-point scales to binary labels in multimodal datasets, may oversimplify the nuanced nature of movie criticism~\citep{socher2013recursive}. Many reviews express mixed sentiments or qualified opinions that are difficult to capture in binary classifications. Professional critics often provide sophisticated analyses that appreciate certain aspects of films while criticizing others, creating sentiment expressions that resist simple categorization.

The binary paradigm also limits evaluation of models' ability to handle the explainability challenge, as it provides limited insight into whether models are making predictions for appropriate reasons~\citep{ribeiro2016should}. A model might achieve high accuracy while relying on spurious correlations or biased patterns in training data, leading to unreliable performance when deployed in different domains or time periods.

\begin{table}[H]
\caption{Common Evaluation Metrics and Their Limitations for Persistent Challenges.}
\label{tab:metrics_limitations}
\centering
\small
\begin{tabularx}{\textwidth}{llXX}
\toprule
\textbf{Metric} & \textbf{Formula/Description} & \textbf{Advantages} & \textbf{Limitations for Challenges} \\
\midrule
Accuracy & $\frac{TP + TN}{TP + TN + FP + FN}$ & Intuitive, widely used & Cannot detect sarcasm failures, domain drift, or bias issues \\
\midrule
Macro-F1 & $\frac{1}{2}(F1_{pos} + F1_{neg})$ & Balanced class performance & Ignores explainability and resource efficiency \\
\midrule
ROC-AUC & Area under ROC curve & Threshold-independent & No insight into robustness across domains/time \\
\midrule
MCC & $\frac{TP \cdot TN - FP \cdot FN}{\sqrt{(TP+FP)(TP+FN)(TN+FP)(TN+FN)}}$ & Handles imbalance well & Missing context processing and efficiency assessment \\
\bottomrule
\end{tabularx}
\end{table}

\subsection{Emerging Recognition of Benchmark Limitations}

\subsubsection{The Saturation Effect and Diminishing Returns}

The progression from approximately 82\% accuracy with bag-of-words approaches to 97--98\% with large language models represents remarkable technical achievement, yet this performance increase has revealed a critical insight: traditional benchmarks may be approaching their utility limits for driving meaningful research progress~\citep{brown2020language,touvron2023llama}. As models achieve near-perfect performance on standard benchmarks, the five persistent challenges become more apparent and more critical for real-world deployment success.

This performance saturation suggests that future progress requires moving beyond accuracy optimization toward holistic approaches that address multiple challenges simultaneously~\citep{rogers2020primer}. The diminishing returns from pure accuracy improvements have redirected research attention toward robustness, interpretability, and practical deployment considerations that existing benchmarks may not adequately capture.

\subsubsection{Domain and Temporal Representation Gaps}

Most benchmark datasets represent snapshots from particular time periods and platforms, potentially limiting their ability to capture temporal drift and domain variation that characterize real-world deployment scenarios~\citep{blitzer2007biographies}. The IMDB-Large dataset, despite its size and influence, reflects language patterns and cultural references from its collection period, which may not generalize to contemporary movie criticism or different platforms.

The focus on English-language reviews from primarily Western cultural contexts also limits benchmark utility for understanding cross-cultural sentiment expression and the challenges of deploying sentiment analysis systems globally~\citep{zhang2018deep}. Movie criticism reflects diverse cultural perspectives and linguistic patterns that may not be captured in existing benchmarks, suggesting need for more diverse and culturally representative evaluation frameworks.

\subsubsection{Inadequate Evaluation of Persistent Challenges}

Traditional benchmarks provide limited assessment of model performance on the five persistent challenges identified in this review. Sarcasm detection capabilities are not systematically evaluated, domain adaptation robustness is not measured across different platforms or time periods, and explainability is not assessed through standard accuracy metrics~\citep{ribeiro2016should,wilson2005recognizing}.

The resource efficiency challenge is particularly underserved by current benchmarks, which typically focus on maximizing accuracy without considering computational constraints, latency requirements, or energy consumption that characterize real-world deployment scenarios~\citep{hu2022lora,dettmers2023qlora}. This gap between benchmark evaluation and deployment reality has led to systems that perform excellently in research settings but struggle in production environments.

\subsection{Toward More Comprehensive Evaluation Frameworks}

\subsubsection{Multi-Dimensional Assessment Needs}

The field's maturation demands evaluation approaches that move beyond traditional accuracy measurements toward comprehensive assessment of model capabilities across multiple dimensions~\citep{rogers2020primer}. Future benchmark development should incorporate explicit evaluation of sarcasm detection, domain adaptation, temporal robustness, explainability, and resource efficiency alongside traditional accuracy metrics.

Temporal robustness testing, where models are evaluated on reviews from different time periods, would provide insights into adaptation capabilities that current benchmarks cannot capture~\citep{blitzer2007biographies}. Cross-domain evaluation using reviews from different platforms and contexts would assess generalization abilities that are critical for real-world deployment but underexplored in current research.

\subsubsection{Adversarial and Stress Testing}

Systematic evaluation of model robustness through adversarial examples and stress testing represents another direction for benchmark evolution~\citep{ribeiro2016should}. Current benchmarks provide limited assessment of how models handle edge cases, deliberate attempts at deception, or inputs that are designed to reveal failure modes.

Sarcasm-specific evaluation metrics that go beyond simple accuracy to assess whether models are identifying sarcastic content for appropriate reasons would provide more meaningful insights into model capabilities~\citep{wilson2005recognizing}. Similarly, explainability metrics that evaluate whether model explanations align with human reasoning about sentiment would be valuable for building trust in deployed systems.

\subsubsection{Resource-Aware Evaluation}

The development of evaluation frameworks that explicitly consider resource constraints represents a critical need for bridging the gap between research and deployment~\citep{hu2022lora}. These frameworks should assess model performance across different computational budgets, enabling direct comparison of accuracy-efficiency trade-offs that are central to real-world decision making.

Latency-aware evaluation, energy consumption measurement, and model size constraints should be incorporated into benchmark standards to ensure that research progress translates to deployable systems~\citep{dettmers2023qlora}. This shift would encourage development of models that balance multiple performance dimensions rather than optimizing accuracy alone.

\subsection{The Path Forward: Integrated Benchmark Design}

The evolution of movie review sentiment analysis benchmarks reflects the field's maturation from simple classification problems to complex real-world deployment challenges~\citep{zhang2018deep,yadav2020sentiment}. While traditional datasets like MR and IMDB-Large have been invaluable for standardizing evaluation and enabling reproducible research, their limitations have become apparent as the field approaches the frontiers defined by the five persistent challenges.

Future benchmark development should embrace integrated evaluation frameworks that assess multiple dimensions of model performance simultaneously, moving beyond accuracy optimization toward holistic system design~\citep{rogers2020primer}. This evolution requires collaboration between academic researchers and industry practitioners to ensure that evaluation standards reflect both theoretical advances and practical deployment realities.

The success of movie review sentiment analysis as a research domain demonstrates the power of well-designed benchmarks to drive sustained progress over decades. As the field continues to evolve, the development of more comprehensive, challenging, and realistic evaluation frameworks will be essential for addressing the persistent challenges that define the next frontier of sentiment analysis research and deployment.

\section{Current Sentiment Analysis Models and Methods}

Sentiment analysis on movie reviews has progressed from lexicon-based heuristics that barely outperformed chance to transformers surpassing 96\% accuracy on the 50k-review IMDb benchmark. Accuracy gains came in four waves lexicons, classic ML, deep neural networks, and large-scale pre-trained transformers each wave improving speed, data efficiency, and interpretability.

\subsection{Rule-Based and Lexicon-Based Methods}

Rule-based and lexicon-based methods form the earliest category of sentiment analysis techniques. These methods operate by leveraging sentiment lexicons precompiled lists of words annotated with polarity values to assess the sentiment of a given text without requiring supervised learning. Despite being conceptually simple and interpretable, these methods face limitations in handling domain-specific expressions, sarcasm, and compositionality.

\subsubsection{SentiWordNet}

SentiWordNet~\citep{baccianella2010sentiwordnet} is a lexical resource derived from WordNet, where each synset (set of synonyms) is assigned three sentiment scores: positivity, negativity, and objectivity. These scores are determined through semi-supervised learning techniques and allow for a graded view of sentiment intensity. In the context of movie reviews, SentiWordNet has been used to compute sentiment scores by aggregating the polarity of terms found in the review text~\citep{esuli2006sentiwordnet}. However, its effectiveness depends heavily on accurate word sense disambiguation a nontrivial challenge, especially in informal or figurative language common in user-generated movie reviews.

\subsubsection{VADER}

VADER (Valence Aware Dictionary and sEntiment Reasoner) is a rule-based sentiment analysis tool specifically tuned for social media and short text contexts~\citep{hutto2014vader}. It combines a human-validated lexicon with grammatical heuristics (e.g., punctuation, capitalization, degree modifiers) to infer sentiment. Unlike traditional lexicon-based approaches, VADER performs well even without prior domain adaptation, making it suitable for lightweight applications such as movie review summarization or preliminary filtering~\citep{hutto2014vader}. Nonetheless, its performance degrades when faced with domain-specific jargon or multi-aspect sentiment common in longer movie reviews.

\subsubsection{LIWC}

The Linguistic Inquiry and Word Count (LIWC) tool is a psycholinguistic lexicon that maps words to psychologically meaningful categories, including affective processes such as positive and negative emotion~\citep{pennebaker2001linguistic}. While not originally designed for sentiment classification, LIWC has been widely applied in the analysis of movie reviews and online discourse to reveal patterns of emotional expression~\citep{tausczik2010psychological}. LIWC's strength lies in its interpretability and grounding in psychological theory, but it lacks coverage for modern slang, idiomatic expressions, and evolving linguistic trends, which limits its applicability in dynamic domains like film critique.

\subsubsection{Advantages and Limitations}

Lexicon-based methods are advantageous due to their interpretability, domain-independence (at least initially), and low computational cost. They are particularly useful in low-resource settings where labeled data is scarce. However, their major limitations include:

\begin{itemize}[leftmargin=*]
    \item \textbf{Lack of domain adaptation:} Lexicons do not capture domain-specific usage (e.g., ``dark'' may be neutral in general language but positive in the context of horror films).
    \item \textbf{Inability to model context:} These methods typically ignore syntactic and semantic context, leading to incorrect sentiment predictions in the presence of negation or sarcasm~\citep{pang2002thumbs}.
    \item \textbf{Static word representations:} Lexicon scores are fixed and do not account for polysemy or word sense variation across contexts.
\end{itemize}

Despite these limitations, rule-based approaches continue to serve as strong baselines and are often integrated with machine learning systems to provide interpretable explanations or initial polarity signals.

\subsection{Traditional Machine Learning Approaches}

Prior to the advent of deep learning, sentiment analysis on movie reviews was predominantly approached as a supervised text classification problem using traditional machine learning algorithms. These methods required careful feature engineering to transform raw text into structured representations, followed by the application of classification algorithms such as Na\"ive Bayes~\citep{teller2000speech}, Support Vector Machines (SVMs)~\citep{https://doi.org/10.1111/j.1467-8640.2006.00277.x}, and Logistic Regression~\citep{10.1145/1871437.1871487}. Though largely surpassed by neural methods in recent years, traditional models remain relevant for their interpretability, speed, and competitive performance on smaller or well-curated datasets.

\subsubsection{Feature Engineering for Sentiment Representation}

One of the foundational steps in traditional sentiment analysis pipelines is the transformation of text into numerical features. The simplest and most common representations include bag-of-words (BoW) and $n$-gram models. Pang et al.~\citep{pang2002thumbs} demonstrated that unigram and bigram features, when combined with frequency-based weighting schemes, can achieve high accuracy on movie reviews.

Term Frequency-Inverse Document Frequency (TF-IDF) is another widely used weighting technique, which downweights commonly occurring words while emphasizing informative terms~\citep{joachims1998text}. TF-IDF has been shown to outperform raw frequency counts by better capturing the discriminative power of specific sentiment-bearing words.

In addition to frequency-based features, part-of-speech (POS) tags have been incorporated to emphasize sentiment-heavy word classes such as adjectives and adverbs. Whitelaw et al.~\citep{whitelaw2005using} proposed enriching BoW models with POS-tagged phrases, leading to improved performance in polarity classification tasks. Feature engineering in traditional pipelines may also include stemming or lemmatization to reduce word form variation, and syntactic parsing to extract structured linguistic patterns.

\subsubsection{Classification Algorithms}

Among the earliest and most effective classifiers for sentiment analysis is the Multinomial Na\"ive Bayes (MNB) algorithm. Despite its simplistic conditional independence assumption, MNB performs competitively on textual data due to its robustness and efficiency~\citep{mccallum1998comparison}. Pang et al.~\citep{pang2002thumbs} found Na\"ive Bayes to perform reasonably well on movie reviews, though it was often outperformed by more discriminative models.

Support Vector Machines (SVMs) quickly became the dominant method for sentiment classification due to their ability to handle high-dimensional sparse data and to find optimal separating hyperplanes in feature space~\citep{joachims1998text}. Linear SVMs, in particular, are well-suited for text classification tasks where data points lie in a high-dimensional feature space. In experiments on the IMDb dataset, SVMs consistently outperformed Na\"ive Bayes and Logistic Regression when paired with appropriate feature selection~\citep{pang2002thumbs}.

Logistic Regression is another frequently used classifier due to its probabilistic output and ability to be interpreted in terms of feature weights. It is often preferred in settings where decision thresholds or confidence scores are required. While generally competitive with SVMs, Logistic Regression may suffer when the classes are not linearly separable, especially in noisy review corpora~\citep{manning2008introduction}.

\subsubsection{Strengths and Limitations}

Traditional machine learning methods offer several advantages. They are computationally efficient and interpretable, allowing for fine control over the modeling pipeline. Feature importance can be directly extracted from model coefficients, making them suitable for use cases that require explainability.

However, these methods suffer from notable limitations. Their performance is heavily dependent on the quality of feature engineering and cannot easily model long-distance dependencies, negation, or compositionality in language. For instance, detecting that ``not a great movie'' carries negative sentiment requires more than just counting positive words. Furthermore, traditional classifiers struggle with domain adaptation and sarcasm detection due to their inability to capture contextual semantics~\citep{pang2002thumbs}.

Despite these drawbacks, traditional models laid the foundational groundwork for sentiment analysis and continue to serve as strong baselines and components in hybrid systems.

\subsection{Deep Learning-Based Methods}

Deep learning has significantly advanced the state of sentiment analysis by enabling end-to-end learning of hierarchical and contextual representations from raw text. Unlike traditional machine learning approaches that rely heavily on manual feature engineering, deep neural networks can automatically learn semantic and syntactic features that are critical for sentiment classification. In the domain of movie reviews, deep learning methods have demonstrated superior performance across multiple benchmarks by capturing compositional sentiment, long-range dependencies, and subtle linguistic cues.

\subsubsection{Convolutional Neural Networks (CNNs)}

Convolutional Neural Networks (CNNs), initially developed for computer vision, have been successfully adapted for text classification tasks such as sentiment analysis. Kim~\citep{kim2014convolutional} showed that a simple CNN with a single convolutional layer applied to pre-trained word vectors (e.g., Word2Vec) can achieve competitive results on multiple sentiment benchmarks, including the IMDb movie review dataset. CNNs are particularly effective at capturing local patterns such as sentiment-bearing phrases (e.g., ``utterly disappointing'', ``brilliant performance'') by applying filters over word sequences. Their parallelizable architecture and relatively shallow depth also make them computationally efficient.

\subsubsection{Recurrent Neural Networks (RNNs), LSTMs, and BiLSTMs}

Recurrent Neural Networks (RNNs) are designed to model sequential dependencies in text, making them well-suited for tasks where word order and context matter. However, traditional RNNs suffer from vanishing and exploding gradient problems, which limit their ability to capture long-term dependencies~\citep{bengio1994learning}. To overcome this, Long Short-Term Memory (LSTM) networks were introduced~\citep{hochreiter1997long}. LSTMs use gating mechanisms to regulate the flow of information, enabling them to remember sentiment-relevant features over longer spans of text.

Bidirectional LSTMs (BiLSTMs) further enhance performance by processing the input sequence in both forward and backward directions, thereby incorporating both past and future context~\citep{tai2015improved}. In sentiment classification of movie reviews, where sentiment may depend on distant contextual cues (e.g., ``Although the film starts slow, it ultimately delivers a powerful message''), BiLSTMs provide a robust mechanism for modeling such dependencies.

\subsubsection{Word Embeddings: Word2Vec and GloVe}

A crucial component of deep learning-based sentiment models is the use of distributed word representations or embeddings. Word2Vec~\citep{mikolov2013distributed}, trained using skip-gram or CBOW objectives, captures semantic similarity by placing similar words in nearby vector space. GloVe~\citep{pennington2014glove}, on the other hand, leverages global co-occurrence statistics to produce embeddings that better capture linear substructures. These embeddings are often used as input to CNNs or RNNs and can be fine-tuned during training for improved task-specific performance. Pre-trained embeddings alleviate data sparsity issues and help models generalize better on small or imbalanced movie review datasets.

\subsubsection{Attention Mechanisms}

While RNNs and CNNs are capable of learning useful features, they often compress long sequences into a single vector, potentially losing important information. Attention mechanisms address this limitation by allowing the model to focus selectively on relevant parts of the input when making predictions~\citep{bahdanau2015neural}. In sentiment analysis, attention helps identify sentiment-bearing phrases or clauses even if they are distant from each other in the input. Yang et al.~\citep{yang2016hierarchical} proposed a Hierarchical Attention Network (HAN) that applies attention at both the word and sentence levels, achieving strong performance on document-level sentiment classification tasks, including movie reviews.

\subsubsection{Summary and Limitations}

Deep learning methods significantly outperform traditional approaches in sentiment classification, particularly on large datasets. They offer the ability to model complex linguistic phenomena such as negation, sarcasm, and compositionality. However, they require substantial computational resources and are often data-hungry. Moreover, deep models are typically less interpretable than their traditional counterparts, making it challenging to understand their decision-making processes without auxiliary explanation methods.

\subsection{Transformer-Based Models}

The introduction of transformer architectures~\citep{vaswani2017attention} has revolutionized the field of natural language processing, including sentiment analysis of movie reviews. Unlike traditional RNN-based models, transformers leverage self-attention mechanisms to model long-range dependencies in text efficiently, without relying on sequential data processing.

\subsubsection{Pretrained Transformer Models}

One of the earliest and most impactful transformer-based models is BERT (Bidirectional Encoder Representations from Transformers)~\citep{devlin2019bert}. BERT is pretrained on large-scale corpora using masked language modeling and next sentence prediction, and can be fine-tuned on downstream tasks such as sentiment classification with minimal architectural modifications. In the context of movie reviews, fine-tuning BERT on datasets like IMDb has yielded state-of-the-art results in binary sentiment classification~\citep{sun2019fine}.

Variants such as RoBERTa~\citep{liu2019roberta} improved upon BERT by optimizing pretraining strategies, including removal of the next sentence prediction objective and training on larger corpora with dynamic masking. These modifications have led to improved sentiment classification performance across benchmarks. DistilBERT~\citep{sanh2019distilbert}, a compressed version of BERT, provides a trade-off between inference speed and accuracy, making it suitable for real-time applications without substantial degradation in accuracy. XLNet~\citep{yang2019xlnet}, a generalized autoregressive pretraining model, overcomes some limitations of BERT by modeling bidirectional context while retaining autoregressive properties, demonstrating competitive performance on sentiment tasks.

\subsubsection{Domain-Specific Transformers}

Although general-purpose pretrained transformers offer strong baselines, domain-adapted transformer models have emerged to capture nuances specific to movie reviews. BERT models for sentiment analysis in different contexts have gained prominence including sentiment analysis for microblogging platforms like Twitter, demonstrating that BERT combined with neural network architectures (CNN, RNN, BiLSTM) achieves superior performance compared to traditional Word2vec approaches~\citep{s23010506}. Batra et al.~\citep{Batra_2021} have applied BERT-based models to software engineering contexts, analyzing sentiment in GitHub comments, Jira comments, and Stack Overflow posts, where ensemble BERT models and compressed BERT variants showed 6--12\% improvement in F1-scores over existing tools like SentiCR and SentiStrength-SE. Penha et al.~\citep{10.1145/3383313.3412249} investigated how much factual knowledge about recommendation items (books, movies, music) is stored in pre-trained BERT models and whether this knowledge can be leveraged for Conversational Recommender Systems (CRS). Their study finds that BERT contains substantial content-based knowledge about items but has limited collaborative knowledge, and while it can perform basic recommendation tasks, it struggles with conversational recommendations when faced with challenging data. More recently, BERT has been combined with BiLSTM for movie review sentiment analysis~\citep{nkhata2025finetuningbertbilstm}, achieving better accuracy than existing state-of-the-art methods and demonstrating how the approach can predict overall movie sentiment for recommendation systems.

\subsubsection{Prompt-Based and Zero-Shot Models}

With the advent of generative pretrained transformers like T5~\citep{raffel2020exploring} and GPT~\citep{brown2020language}, few-shot and zero-shot sentiment classification has become increasingly feasible. These models can be prompted using task-specific instructions, enabling them to perform classification tasks without extensive fine-tuning. For example, a prompt like ``Classify the sentiment of this review: `The movie was breathtaking and unforgettable.''' can elicit accurate sentiment predictions even with minimal supervision. However, prompt-based performance is often sensitive to prompt wording and may require careful calibration~\citep{gao-etal-2021-making}.

\subsubsection{Strengths and Limitations}

Transformer-based models offer several advantages for sentiment classification, including superior accuracy, transferability to low-resource settings, and the ability to model complex syntactic and semantic relationships. However, they are not without limitations. Large models such as BERT and GPT-3 are computationally expensive to train and deploy, and their performance may degrade when faced with domain-specific jargon or informal language unless further fine-tuned. Additionally, explainability remains a significant challenge, especially in high-stakes applications where interpretability of predictions is critical.

\subsection{Large Language Models and the New Paradigm}

The emergence of Large Language Models (LLMs) such as GPT-4, Llama, and Claude has fundamentally transformed the landscape of sentiment analysis, introducing capabilities that extend far beyond traditional fine-tuning paradigms.

\subsubsection{Zero-Shot and Few-Shot Sentiment Classification}

A comprehensive evaluation by Zhang et al.~\citep{zhang2024sentiment} across 13 sentiment analysis tasks on 26 datasets revealed that while LLMs demonstrate satisfactory performance in simpler sentiment classification tasks, they lag behind fine-tuned small language models (SLMs) in more complex tasks requiring deeper understanding of specific sentiment phenomena. However, LLMs significantly outperform SLMs in few-shot learning settings, suggesting their potential when annotation resources are limited. The study introduced SentiEval, a benchmark for more comprehensive evaluation of LLM sentiment capabilities.

Krugmann and Hartmann~\citep{krugmann2024sentiment} benchmarked GPT-3.5, GPT-4, and Llama 2 against established transfer learning models for marketing research applications. Their findings indicate that despite their zero-shot nature, LLMs can not only compete with but in some cases surpass traditional transfer learning methods in sentiment classification accuracy. The study also highlighted that prompt specificity significantly affects prediction consistency, with explicitly structured prompts yielding greater reliability.

Recent work on multilingual sentiment analysis has demonstrated GPT's effectiveness across 12 languages, with performance comparable to or exceeding top-performing fine-tuned models from previous years~\citep{pnas2024gpt}. GPT-4 achieved correlations of $r = 0.59$ to $0.77$ with human annotators, substantially outperforming traditional dictionary-based methods ($r = 0.20$ to $0.30$).

\subsubsection{Chain-of-Thought Prompting for Sentiment}

Chain-of-thought (CoT) prompting has emerged as a powerful technique for improving LLM performance on sentiment tasks requiring reasoning. Fei et al.~\citep{fei2023reasoning} introduced the Three-hop Reasoning (THOR) framework for implicit sentiment analysis, which mimics human-like reasoning by step-by-step inducing the implicit aspect, opinion, and finally the sentiment polarity. This approach addresses the challenge of detecting sentiment where opinion cues are implicit and obscure, requiring common-sense and multi-hop reasoning to infer latent intent.

The THOR framework demonstrates that structured reasoning chains can significantly improve performance on nuanced sentiment tasks, particularly for sarcasm and implicit sentiment where traditional classification approaches struggle. This represents a paradigm shift from pattern matching to genuine reasoning about sentiment.

\subsubsection{Instruction-Tuned Models for Aspect-Based Sentiment}

InstructABSA~\citep{scaria2024instructabsa} introduced an instruction learning paradigm for Aspect-Based Sentiment Analysis (ABSA) that achieves state-of-the-art results across multiple subtasks. By incorporating positive, negative, and neutral examples into training samples and instruction-tuning the Tk-Instruct model, InstructABSA outperformed previous SOTA on the Rest14 ATE subtask by 5.69\% points and the Rest15 ATSC subtask by 9.59\% points, surpassing models 7$\times$ larger. Notably, just 50\% of training data was sufficient to achieve competitive results, demonstrating remarkable sample efficiency.

\subsubsection{Limitations and Emerging Challenges}

Despite their impressive capabilities, LLMs present new challenges for sentiment analysis. Model outputs exhibit sensitivity to prompt wording, with even small changes in phrasing leading to different sentiment classifications~\citep{zhang2024sentiment}. Reproducibility remains a concern, as temperature settings and model versions can affect result consistency. Furthermore, the computational cost of deploying LLMs at scale poses practical limitations for real-time sentiment monitoring applications.

The question of whether LLMs truly ``understand'' sentiment or merely exploit statistical patterns remains open. While chain-of-thought prompting improves interpretability, the black-box nature of these models continues to challenge deployment in high-stakes domains requiring explainable predictions.

\begin{figure}[H]
\centering
\begin{tikzpicture}[scale=0.8]
\draw[thick,->] (0,0) -- (12,0) node[right] {Year};
\draw[thick,->] (0,0) -- (0,8) node[above] {Accuracy (\%)};

\foreach \y/\label in {1/80, 2/85, 3/90, 4/95, 5/100} {
    \draw (-0.1,\y) -- (0.1,\y);
    \node[left] at (-0.2,\y) {\footnotesize \label};
}

\foreach \x/\year in {2/2005, 4/2010, 6/2015, 8/2020, 10/2025} {
    \draw (\x,-0.1) -- (\x,0.1);
    \node[below] at (\x,-0.2) {\footnotesize \year};
}

\draw[thick, blue] plot[smooth] coordinates {
    (1,1.2) (3,1.8) (5,2.8) (7,4.2) (9,4.7) (10,4.8)
};

\draw[thick, red, dashed] plot[smooth] coordinates {
    (1,0.5) (3,0.7) (5,1.2) (7,2.5) (9,4.0) (10,5.5)
};

\fill[gray!20] (8.5,4) rectangle (11,5.5);
\node[text width=2.5cm, align=center] at (12.75,6) {\footnotesize \textit{Performance}\\ \textit{Saturation Zone}};

\node[right] at (1,7) {\footnotesize \textcolor{blue}{ } Peak Accuracy};
\node[right] at (1,6.5) {\footnotesize \textcolor{red}{- -} Challenge Awareness};

\node[text width=2cm, align=center] at (2,2.5) {\footnotesize Bag-of-Words\\ 82\%};
\node[text width=2cm, align=center] at (5,3.5) {\footnotesize RNNs\\ 90--93\%};
\node[text width=2cm, align=center] at (7.5,5) {\footnotesize Transformers\\ 94--97\%};
\node[text width=2cm, align=center] at (10,5.5) {\footnotesize LLMs\\ 97--98\%};

\end{tikzpicture}
\caption{The progression of peak accuracy on IMDB benchmarks versus growing awareness of persistent challenges. As accuracy approaches saturation, attention shifts to robustness, explainability, and deployment considerations.}
\label{fig:accuracy_vs_challenges}
\end{figure}
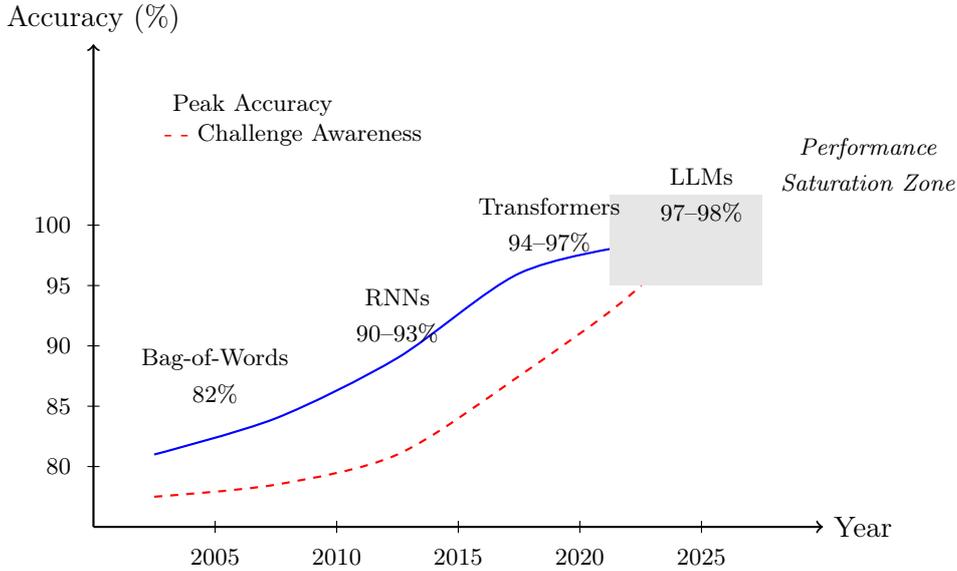

\section{Domain-Specific Challenges}

Despite the significant advancements in sentiment analysis, a number of domain-specific challenges continue to hinder the accuracy and robustness of models. These challenges are particularly pronounced in real-world applications such as e-commerce, movie reviews, and social media analytics, where language use is highly dynamic and context-dependent.

\subsection{Sarcasm and Irony Detection}

One of the most persistent challenges in sentiment classification is the accurate detection of sarcasm and irony. Sarcasm often conveys a sentiment opposite to the literal meaning of the text, making traditional sentiment classifiers fail when relying on lexical cues alone. For example, the statement \textit{``Great, another three-hour delay!''} expresses negative sentiment despite containing the positive word ``great.'' Studies have shown that sarcasm significantly reduces classification accuracy when left unaddressed~\citep{10.1145/3124420}. Recent work leverages contextual embeddings and attention mechanisms to better capture incongruity between expected and actual sentiment~\citep{ghosh-etal-2015-semeval}. However, sarcasm detection remains challenging due to its dependence on pragmatic cues, cultural background, and shared speaker-listener knowledge~\citep{kreuz1996sarcasm}.

\subsection{Domain Drift}

Domain drift refers to the deterioration in model performance when applying sentiment classifiers trained on one domain to another. For instance, models trained on product reviews may not generalize effectively to movie or restaurant reviews due to domain-specific vocabulary and context~\citep{blitzer2007biographies}. Domain adaptation techniques, including adversarial learning and pivot-based feature transfer, have been proposed to mitigate this issue~\citep{glorot2011domain}. Nevertheless, achieving robustness across domains is particularly challenging when labeled data in the target domain is scarce.

\subsection{Temporal Drift}

Language evolves over time, and with it, the expressions of sentiment also change. Temporal drift refers to the decline in model accuracy as a result of evolving slang, memes, and cultural references~\citep{diaz2016query}. For example, terms like ``sick'' or ``fire'' may shift from negative to positive sentiment over time. Research has demonstrated that models trained on static data often underperform on more recent corpora~\citep{10.5555/3540261.3542508}. Addressing this requires continual learning approaches that update model parameters in response to new linguistic trends, though such solutions often face the risk of catastrophic forgetting as first explained in the seminal 1989 paper by McCloskey et al.~\citep{ffe0793b43f842d2a50467d736a80c83}.

\subsection{Long-form Context}

Another key challenge is sentiment analysis over long-form content such as blogs, articles, and movie scripts. Unlike short reviews or tweets, longer texts often contain mixed sentiments across different sections, requiring hierarchical models that capture local and global dependencies~\citep{yang2016hierarchical}. Models such as Hierarchical Attention Networks (HANs) have been shown to improve performance by aggregating sentence-level embeddings into document-level representations~\citep{yang2016hierarchical}. However, accurately modeling sentiment across long documents remains computationally intensive and often struggles with discourse-level phenomena such as contrastive statements and sentiment shifts within the same text~\citep{bhatia-etal-2015-better}.

\subsection{Cultural and Language Diversity}

Sentiment expression varies widely across languages and cultures. Words, idioms, and symbols that carry strong sentiment in one culture may not translate equivalently into another~\citep{banea-etal-2008-multilingual}. For instance, the interpretation of emojis varies significantly between cultural contexts~\citep{10.1371/journal.pone.0144296}. Cross-lingual sentiment analysis approaches, including multilingual embeddings and machine translation-based methods, attempt to address this challenge~\citep{barnes-etal-2018-bilingual}. However, even state-of-the-art multilingual transformers such as XLM-R struggle with low-resource languages and culture-specific idiomatic expressions~\citep{conneau-etal-2020-unsupervised}. This diversity highlights the need for culturally adaptive sentiment models that integrate both linguistic and socio-cultural knowledge.

In practice, these domain-specific challenges rarely occur in isolation. Sarcasm may intertwine with cultural idioms, temporal drift may alter how irony is expressed, and domain drift may exacerbate the failure to detect nuanced sentiment in multimodal reviews. For instance, a sarcastic movie review written in Hinglish (mixture of Hindi and English) on YouTube in 2023 might be nearly impossible to classify correctly using a model trained on English-only IMDB reviews from 2010. This interplay underscores the necessity for holistic approaches that combine sarcasm detection, domain adaptation, temporal robustness, and multilingual modeling into unified frameworks~\citep{barnes-etal-2018-bilingual}. Addressing these intertwined challenges is a key direction for future research in sentiment analysis of movie reviews.

\section{Future Research Directions}

The field of sentiment analysis on movie reviews continues to evolve rapidly, with several emerging research challenges and opportunities shaping its future trajectory. In this section, we discuss key areas that demand attention, including improved benchmarks, few-shot and zero-shot learning, cross-lingual and cross-domain transfer, explainability, multimodal extensions, and fairness in sentiment analysis.

\subsection{Improving Benchmarks}

Although numerous benchmark datasets exist for sentiment analysis, they often fail to capture nuanced linguistic phenomena such as sarcasm, irony, or evolving cultural expressions~\citep{gonzalez-ibanez-etal-2011-identifying, 10.1145/3124420}. Benchmark datasets also rarely incorporate domain and temporal drift~\citep{maas2011learning}, which limits the generalizability of models when exposed to changing review trends and movie discourse. Future benchmarks need to incorporate long-form contextual reviews~\citep{beltagy2020longformerlongdocumenttransformer} and scenarios that require deeper discourse understanding rather than sentence-level polarity predictions. Moreover, evaluation frameworks should integrate dimensions of explainability~\citep{danilevsky-etal-2020-survey} and computational efficiency~\citep{schwartz2020green} to ensure real-world applicability in resource-constrained environments.

\begin{table}[H]
\caption{Proposed Requirements for Next-Generation Movie Review Sentiment Analysis Benchmarks.}
\label{tab:future_benchmarks}
\centering
\small
\begin{tabularx}{\textwidth}{lXX}
\toprule
\textbf{Challenge} & \textbf{Current Gap} & \textbf{Proposed Evaluation} \\
\midrule
Sarcasm Detection & No systematic evaluation of sarcastic content & Dedicated sarcasm subset with explanation requirements \\
\midrule
Domain Drift & Single platform/source evaluation only & Multi-platform evaluation with cross-domain generalization \\
\midrule
Temporal Drift & Static datasets from specific time periods & Temporal robustness testing across multiple time periods \\
\midrule
Long-form Context & Truncation or chunking of longer reviews & Full document processing evaluation metrics \\
\midrule
Explainability & No explanation assessment in standard metrics & Human-aligned explanation evaluation frameworks \\
\midrule
Resource Efficiency & Accuracy optimization without cost consideration & Pareto frontier evaluation (accuracy vs. efficiency) \\
\bottomrule
\end{tabularx}
\end{table}

\subsection{Beyond Few-shot: Emerging LLM Challenges}

While few-shot and zero-shot learning approaches have rapidly matured with the advent of large language models~\citep{brown2020language,zhang2024sentiment,yin-etal-2019-benchmarking}, new challenges have emerged. Current LLMs demonstrate strong performance on standard sentiment benchmarks but struggle with complex tasks requiring deeper understanding of specific sentiment phenomena~\citep{zhang2024sentiment}. Key open problems include: (1) reducing sensitivity to prompt wording, where minor phrasing changes can dramatically alter predictions~\citep{krugmann2024sentiment}; (2) ensuring reproducibility across model versions and temperature settings; (3) developing efficient inference strategies for real-time applications given the computational cost of LLM deployment; and (4) addressing the gap between benchmark performance and real-world robustness on domain-specific jargon, evolving language, and adversarial inputs. Chain-of-thought prompting~\citep{fei2023reasoning,wei2022chain} and prompt tuning approaches~\citep{gu-etal-2022-ppt} offer promising directions for improving reasoning on implicit sentiment, but systematic evaluation frameworks for such approaches remain underdeveloped.

\subsection{Cross-lingual and Cross-domain Transfer}

Movie reviews are inherently multilingual and cross-cultural, yet much of the sentiment analysis research has focused on English-only datasets, like the seminal product review paper by Hu et al.~\citep{10.1145/1014052.1014073}. Cross-lingual transfer learning using multilingual models such as XLM-R~\citep{conneau-etal-2020-unsupervised} or mBERT~\citep{devlin2019bert} offers promising directions for extending sentiment analysis systems to underrepresented languages. Furthermore, cross-domain transfer is essential for adapting models trained on professional critic reviews to informal user reviews or social media commentary~\citep{glorot2011domain}. Addressing these challenges requires robust domain adaptation and multilingual embedding techniques to ensure inclusivity and applicability across diverse populations.

\subsection{Explainable Sentiment Models}

Explainability remains a crucial area of research for sentiment analysis, especially in high-stakes applications such as content recommendation or censorship~\citep{danilevsky-etal-2020-survey}. While attention-based mechanisms~\citep{vaswani2017attention} and post-hoc interpretability tools~\citep{ribeiro2016should} have been explored, they often provide incomplete or misleading explanations. Future research must focus on building inherently interpretable sentiment models that allow end-users to understand why a particular review is classified as positive, negative, or mixed. Such approaches would also enable the identification of biases in training data and the reduction of spurious correlations.

\subsection{Multimodal and Conversational Sentiment}

As user-generated reviews increasingly take multimodal forms, incorporating audio, visual, and textual signals into sentiment analysis has become essential~\citep{zadeh2017tensor, bagher-zadeh-etal-2018-multimodal,lai2023multimodalsentimentanalysissurvey}. Multimodal transformers such as MMBERT~\citep{9434063} represent promising directions for fusing diverse modalities. Additionally, conversational sentiment analysis, where reviewers discuss movies in interactive settings (e.g., podcasts, live streams, or interviews), demands models that can track sentiment across multiple speakers and temporal contexts~\citep{8764449}. These directions open new opportunities for richer, context-aware sentiment understanding.

\subsection{Bias and Fairness in Sentiment Analysis Systems}

Bias and fairness present pressing ethical challenges in sentiment analysis~\citep{sheng-etal-2019-woman}. Models trained on imbalanced review datasets may propagate stereotypes, amplify biases, or underperform on minority languages and cultural expressions. For instance, sentiment polarity markers may vary across dialects or sociolects~\citep{kiritchenko-mohammad-2018-examining}, leading to systematic misclassification. Future research must focus on fairness-aware learning frameworks, debiasing strategies, and representative datasets to ensure equitable performance across demographic and linguistic groups.

\section{Conclusion}

The domain of movie reviews has consistently served as a proving ground for advances in sentiment analysis, pushing the boundaries of natural language understanding. Since the seminal work by Pang et al.~\citep{pang2002thumbs}, movie review datasets have acted as the canonical benchmark for testing supervised and unsupervised methods, ranging from early bag-of-words and SVM classifiers~\citep{pang2004sentimental,turney2002thumbs} to modern pre-trained transformers, multimodal models, and large language models~\citep{devlin2019bert,pmlr-v139-radford21a,liu2019roberta,zhang2024sentiment}. The diversity and richness of movie reviews ranging from concise star ratings to nuanced long-form critiques have forced researchers to grapple with fundamental challenges such as sarcasm, irony, domain drift, temporal shifts, and cultural variation~\citep{gonzalez-ibanez-etal-2011-identifying,8675939,amir-etal-2016-modelling,10.1145/3124420}. These challenges not only stress-test algorithms but also advance the state of the art in natural language processing more broadly.

The emergence of large language models has fundamentally transformed the sentiment analysis landscape. While LLMs demonstrate impressive zero-shot and few-shot capabilities that rival or exceed fine-tuned models on standard benchmarks~\citep{krugmann2024sentiment,pnas2024gpt}, they also introduce new challenges around prompt sensitivity, reproducibility, and computational cost. Chain-of-thought prompting~\citep{fei2023reasoning} and instruction-tuned models~\citep{scaria2024instructabsa} represent promising directions for improving reasoning on nuanced sentiment tasks, particularly for implicit sentiment and aspect-based analysis. However, comprehensive evaluation frameworks like SentiEval~\citep{zhang2024sentiment} reveal that LLMs still lag behind specialized models on complex structured sentiment tasks, suggesting that the field has not yet reached a definitive solution.

Looking ahead, sentiment understanding in the context of creative media remains uniquely positioned to influence the trajectory of research. The development of fine-grained emotion datasets like GoEmotions~\citep{demszky2020goemotions} and unified social media benchmarks like TweetEval~\citep{barbieri2020tweeteval} reflects a broader recognition that binary sentiment classification is insufficient for real-world applications. Cross-lingual transfer learning~\citep{conneau-etal-2020-unsupervised,ranasinghe-zampieri-2020-multilingual} promises to extend sentiment systems to under-resourced languages, while research in explainability~\citep{danilevsky-etal-2020-survey,rajani-etal-2019-explain} and bias mitigation~\citep{kiritchenko-mohammad-2018-examining,sheng-etal-2019-woman} is critical for ensuring that sentiment models remain fair and interpretable.

Moreover, as multimodal and conversational sentiment analysis continues to evolve~\citep{poria2017review,bagher-zadeh-etal-2018-multimodal,Cai2025}, the movie domain will provide a natural testbed for models that must integrate text, audio, and visual signals to capture audience reactions or understand narrative tone. Such advancements have implications that go beyond recommendation systems or review aggregation, influencing how we design human-centered AI capable of interpreting creativity, storytelling, and emotional resonance.

From an industry perspective, the methodological advances discussed in this survey have practical implications for recommendation systems, content moderation, market research, and audience analytics within the film and streaming ecosystem. Fine-grained sentiment analysis can support more personalized content recommendations, early detection of audience dissatisfaction, and improved understanding of viewer engagement beyond coarse rating scores. The ability of modern LLMs to provide explanatory reasoning for their predictions~\citep{fei2023reasoning} opens new possibilities for interpretable audience analytics.

In summary, the movie domain has shaped sentiment analysis into a rich interdisciplinary field that continually challenges researchers to develop methods that are robust, context-aware, and adaptable across cultures and modalities. As sentiment understanding becomes more deeply integrated into creative industries, it will not only enhance our ability to process media at scale, but also provide insights into the ways humans communicate and experience emotion. Future research must balance methodological innovation with cultural sensitivity, explainability, and fairness, ensuring that sentiment analysis contributes meaningfully to our understanding of human expression in creative media.

\section*{Author Contributions}

A.G., S.D. and K.T. have equally contributed to conceptualization, methodology, validation and writing original draft preparation. R.A.Z., A.W.M., and S.J.M. contributed to review, editing, and supervision. All authors have read and agreed to the published version of the manuscript.

\section*{Funding}

This research received no external funding.

\section*{Data Availability}

Not applicable.

\section*{Conflicts of Interest}

The authors declare no conflicts of interest.

\bibliographystyle{ieeetr}
\bibliography{reference}

\end{document}